\documentclass[pdflatex,sn-mathphys]{sn-jnl}

\jyear{2023}%

\theoremstyle{thmstyleone}%
\theoremstyle{thmstyletwo}%

\theoremstyle{thmstylethree}%

\begin{document}

\title[EUV Reflection Ptychography]{Wavelength-multiplexed Multi-mode EUV Reflection Ptychography based on Automatic-Differentiation}


\author*[1,2]{\fnm{Yifeng} \sur{Shao}}\email{Y.Shao@tudelft.nl}
\equalcont{These authors contributed equally to this work.}

\author[2]{\fnm{Sven} \sur{Weerdenburg}}\email{S.Weerdenburg@tudelft.nl}
\equalcont{These authors contributed equally to this work.}

\author[1]{\fnm{Jacob} \sur{Seifert}}\email{J.Seifert@uu.nl}
\equalcont{These authors contributed equally to this work.}

\author[2]{\fnm{H. Paul} \sur{Urbach}}

\author[1]{\fnm{Allard P.} \sur{Mosk}}

\author[2,3]{\fnm{Wim} \sur{Coene}}

\affil[1]{\orgdiv{Nanophotonics}, \orgname{Debye Institute for Nanomaterials Science and Center for Extreme Matter and Emergent Phenomena}, \state{Utrecht University}, \orgaddress{\street{P.O. Box 80000}, \city{Utrecht}, \postcode{3508 TA}, \country{The Netherlands}}}

\affil[2]{\orgdiv{Imaging Physics Department}, \orgname{Applied Science Faculty}, \state{Delft University of Technology}, \orgaddress{\street{Lorentzweg 1}, \city{Delft}, \postcode{2628 CJ }, \country{The Netherlands}}}

\affil[3]{\orgdiv{Research Department}, \orgname{ASML Netherlands B.V}, \orgaddress{\street{De Run 6501}, \city{Veldhoven}, \postcode{5504 DR }, \country{The Netherlands}}}


\abstract{Ptychographic extreme ultraviolet (EUV) diffractive imaging has emerged as a promising candidate for the next-generation metrology solutions in the semiconductor industry, as it can image wafer samples in reflection geometry at the nanoscale. This technique has surged attention recently, owing to the significant progress in high-harmonic generation (HHG) EUV sources and advancements in both hardware and software for computation.

In this study, a novel algorithm is introduced and tested, which enables wavelength-multiplexed reconstruction that enhances the measurement throughput and introduces data diversity, allowing the accurate characterisation of sample structures. To tackle the inherent instabilities of the HHG source, a modal approach was adopted, which represents the cross-density function of the illumination by a series of mutually incoherent and independent spatial modes.

The proposed algorithm was implemented on a mainstream machine learning platform, which leverages automatic differentiation to manage the drastic growth in model complexity and expedites the computation using GPU acceleration. By optimising over 200 million parameters, we demonstrate the algorithm's capacity to accommodate experimental uncertainties and achieve a resolution approaching the diffraction limit in reflection geometry. The reconstruction of wafer samples with 20-nm heigh patterned gold structures on a silicon substrate highlights our ability to handle complex physical interrelations involving a multitude of parameters. These results establish ptychography as an efficient and accurate metrology tool.}




\maketitle


As the semiconductor industry advances along the trajectory projected by Moore’s law and approaches the ability to manufacture 3D transistors with features on the order of a few nanometers, conventional metrology solutions are confronted with challenges in addressing the shrinking dimensions when characterising the critical dimensions and the overlay errors of the structures printed on the wafer by lithography \cite{mack2011fifty, orji2018metrology}. Due to its capability to perform non-invasive inspections of non-isolated and non-periodic patterns, ptychography emerges as a promising candidate for next-generation metrology solutions \cite{holler2017high, holler2019three, eschen2022material, zhang2015high, porter2017general, tanksalvala2021nondestructive, gardner2017subwavelength, wang2023high}, offering both amplitude and phase information by means of diffractive imaging with a superior resolution and a larger penetration depth into the materials through the use of extreme ultraviolet (EUV) illumination. As a computational approach, it does not require expensive imaging optics, whose components can be extremely challenging to fabricate for wavelengths in the EUV regime \cite{goldberg2015new, benk2015demonstration}. Instead, it illuminates the sample at overlapping areas during a scan process and reconstructs the image of the sample along with the illumination field from the diffraction patterns measured at the corresponding scanning positions via iterative optimisation \cite{rodenburg2007hard, thibault2008high, loetgering2022advances}. 

Compared to the enormous synchrotron or free-electron laser facilities, the compact-sized table-top sources based on high-harmonic generation (HHG) are ideal for industrial applications by providing a spatially moderately coherent illumination consisting of spectral harmonics in the EUV regime \cite{ditmire1996spatial, zerne1997phase, bartels2002generation}. Owing to the overlap between the illuminated sample areas at adjacent scanning positions, the dataset offers abundant information with the necessary redundancy for phase retrieval of illumination and sample fields \cite{guizar2008phase, maiden2009improved}, allowing one to utilise the advantage of the multiple spectral harmonics generated by the HHG source with wavelength-multiplexed reconstruction \cite{batey2014information, zhang2016ptychographic,brooks2022temporal,loetgering2021tailoring}. In the meantime, one can also tackle the inherent instabilities of the HHG source with ptychography using a modal approach, in which one reconstructs a series of mutually incoherent and independent spatial modes of the illumination instead of a single illumination field \cite{whitehead2009diffractive, thibault2013reconstructing, clark2014dynamic, chen2020mixed}. 

Recent progress on HHG sources, especially the significant improvement in brightness and photon flux \cite{tschernajew2020high, kirsche2023continuously}, enables data acquisition to be completed in a reasonable timeframe and has brought EUV ptychography into the spotlight as a potentially viable technology. However, exploiting information redundancy in the dataset with ptychography leads to a drastic growth in model complexity. To manage the entangled relations between diverse physical processes, automatic differentiation (AD) \cite{baydin2018automatic} manifests as an ideal tool for building a universal framework for ptychography models \cite{jurling2014applications, kandel2019using, seifert2021efficient, kharitonov2021flexible, maathuis2022sensor} as well as other computational imaging modalities \cite{du2020three, du2021adorym}, releasing researchers from the laborious work of manually deriving the formulas and implementing the routines for computing various gradients. 

In view of metrology applications in the semiconductor industry, it is desired to use a beamline in reflection instead of transmission geometries \cite{eschen2022material,gardner2017subwavelength, wang2023high, loetgering2021tailoring} for wafer samples. Additionally, extending from single wavelength \cite{zhang2015high, porter2017general, tanksalvala2021nondestructive} to multiple wavelengths can enhance the measurement throughput and introduce data diversity, allowing accurate characterisation of the structure information. Previous works of wavelength-multiplexed reconstruction with HHG sources, however,\cite{zhang2016ptychographic, loetgering2021tailoring} did not incorporate the spatial modes and could not correct experimental uncertainties. Involving all variables in the optimisation is essential to reconstruct the probe and object reliably. 

In this work, we present results obtained on our EUV beamline with an HHG source for ptychographic diffractive imaging in reflection geometry. We are the first to demonstrate that by combining AD with a modular design of the ptychography model, we can achieve wavelength-multiplexed multi-mode reconstruction of both a dispersive sample and the illumination while accounting for various uncertainties that arise during the experiment. Our reconstruction uses no prior knowledge of the object and probe. By leveraging GPU acceleration, we can efficiently explore the full solution space of the inverse problem by jointly optimising over 200 million parameters. 

\subsection{Universal Ptychography Algorithm based on Automatic Differentiation}\label{Software}

Consider the universal case of the ptychography model when the illumination consists of multiple spectral harmonics. Both the field illuminating the sample and the sample itself exhibit moderate spatial partially coherent behaviours, which can be described by multiple spatial modes \cite{wolf1982new, LAHIRI2016605, whitehead2009diffractive, thibault2013reconstructing, clark2014dynamic, chen2020mixed}. The illumination modes represent a decoherence effect as a consequence of the accumulated instabilities, such as source fluctuations or medium turbulences in the gas jet, during the data acquisition time, while the sample modes typically characterise a stationary stochastic process of sample dynamics \cite{wolf1982new, LAHIRI2016605, thibault2013reconstructing}. 

By incorporating the correction of field propagation between tilted planes into the data preprocessing procedure, the transmission and reflection geometries can be treated equally. Our universal ptychography model is based on the following analytical formula to predict the diffraction patterns measured during the scan:

\begin{equation}
\label{eq: simulation}
  I_k(\pmb\rho) = \mathcal{M}\left[\sum^L_{l=1}\sum^M_{m=1}\sum^N_{n=1}\lvert \mathcal{D}_{\lambda_l, z}\left[\sigma_k P_{l,m}(\mathbf{r})O_{l,n}(\mathbf{r}-\mathbf{s}_k)\right](\pmb\rho)\rvert^2+\mathcal{N}(\pmb\rho)\right],
\end{equation}
where $k$ denotes the scan index, $l$ is the index of $L$ wavelengths, $m$ and $n$ are the indices of $M$ illumination modes and $N$ sample modes, respectively. $\mathcal{M}$ represents the measurement-related effects at the detector, such as masking or clipping values of pixels in the presence of saturation and variation of responsivities of pixels (the "hot" and "cold" pixels). $\mathbf{r}$ and $\pmb\rho$ are 2-dimensional vectors representing the sample plane and the camera plane coordinates, respectively. $\mathcal{N}(\pmb\rho)$ denotes the intensity of a common background signal due to spurious reflections in the beamline and specular reflection by the sample reaching the detector. $\mathcal{D}_{\lambda,z}$ represents the propagator that depends on the wavelength $\lambda$ and propagation distance $z$. At wavelength $\lambda_l$, we denote the $m$th illumination mode of the probe by $P_{l,m}(\mathbf{r})$ and the $n$th sample mode of the object by $O_{l,n}(\mathbf{r})$. We further denote the object's scanning position by $\mathbf{s}_k$ and use $\sigma_k$ as a coefficient to scale the probe power for the $k$th acquisition. Using the power coefficient to deal with the energy fluctuation of the HHG source, assuming that the probe shape remains unchanged through the scanning process, is more efficient in memory and computation as compared with the probe relaxation approach \cite{odstrcil2016ptychographic}. The ptychography process and the related experimental setup are shown in Fig.~\ref{fig:computation_graph} (A). 

The difference between the predicted and measured diffraction patterns, denoted by $I_k$ and $\hat{I}_k$, is quantified by the loss function $\mathcal{L}(I_k,\hat{I}_k)$. It is usually defined as the standard mean-squared-error (MSE) on the modulus of the diffracted field. If the noise statistics is known \textit{a priori}, we can also use the maximum-likelihood estimation (MLE) for the probability distribution of the noise, e.g., the Gaussian and Poisson distributions \cite{thibault2012maximum, godard2012noise,odstrvcil2018iterative}. Both MSE and MLE with mixed noise statistics \cite{seifert2023maximum} have been implemented for the loss functions in combination with the Adam optimiser \cite{kingma2014adam}. 

The conventional way of computing the gradient by implementing analytical formulas derived from Eq.~\ref{eq: simulation} severely limits the flexibility of ptychography, because all laborious works, i.e., deriving and implementing the gradient, must be repeated even for the smallest adaptations to the universal model Eq.~\ref{eq: simulation}. Using automatic differentiation (AD) for gradient computation \cite{baydin2018automatic} is a remedy to this challenge. We can interpret the universal model that comprises concatenated sub-models as a composite function with two inputs: $\sigma_k$ and $\mathbf{s}_k$, and one output: the diffraction pattern $I_k$, both specified for a particular scan index $k$. Eq.~\ref{eq: simulation} can be written as:

\begin{equation}
\label{eq: simulation graph}
\begin{aligned}
	  I_k &= \mathbf{f}(\sigma_k, \mathbf{s}_k) =\mathbf{f}_5^{\mathcal{M},\mathcal{N}}\left(\mathbf{f}^{\lambda, z}_4\Big(\mathbf{f}_3\Big(\mathbf{f}^{\mathbf{P}}_1\left(\sigma_k\right), \mathbf{f}_2^{\mathbf{O}}\left(\mathbf{s}_k\right)\Big)\Big)\right),
\end{aligned}
\end{equation}
where the superscripts represent the variables that the respective functions depend on. $\mathbf{P}$ and $\mathbf{O}$ are both 4D data arrays, representing the 2D distributions of all spatial modes at all wavelengths. We define that $\mathbf{f}_1^{\mathbf{P}}$ scales the power of probe $\mathbf{P}$ by $\sigma_k$ and $\mathbf{f}^{\mathbf{O}}_2$ shifts the location of sample $\mathbf{O}$ by $\mathbf{s}_k$. $\mathbf{f}_3$ computes the interaction between the probe and the object, and subsequently, $\mathbf{f}_4^{\lambda, z}$ propagates the resulting field to the detector plane. Finally, $\mathbf{f}_5^{\mathcal{M},\mathcal{N}}$ applies the measurement-related effects to the integrated power of the resulting field on the camera, predicting the diffraction pattern to be measured. For more details on the functions, please refer to the Supplementary Information.  

\begin{figure}[hbt]
  \includegraphics[width=\textwidth]{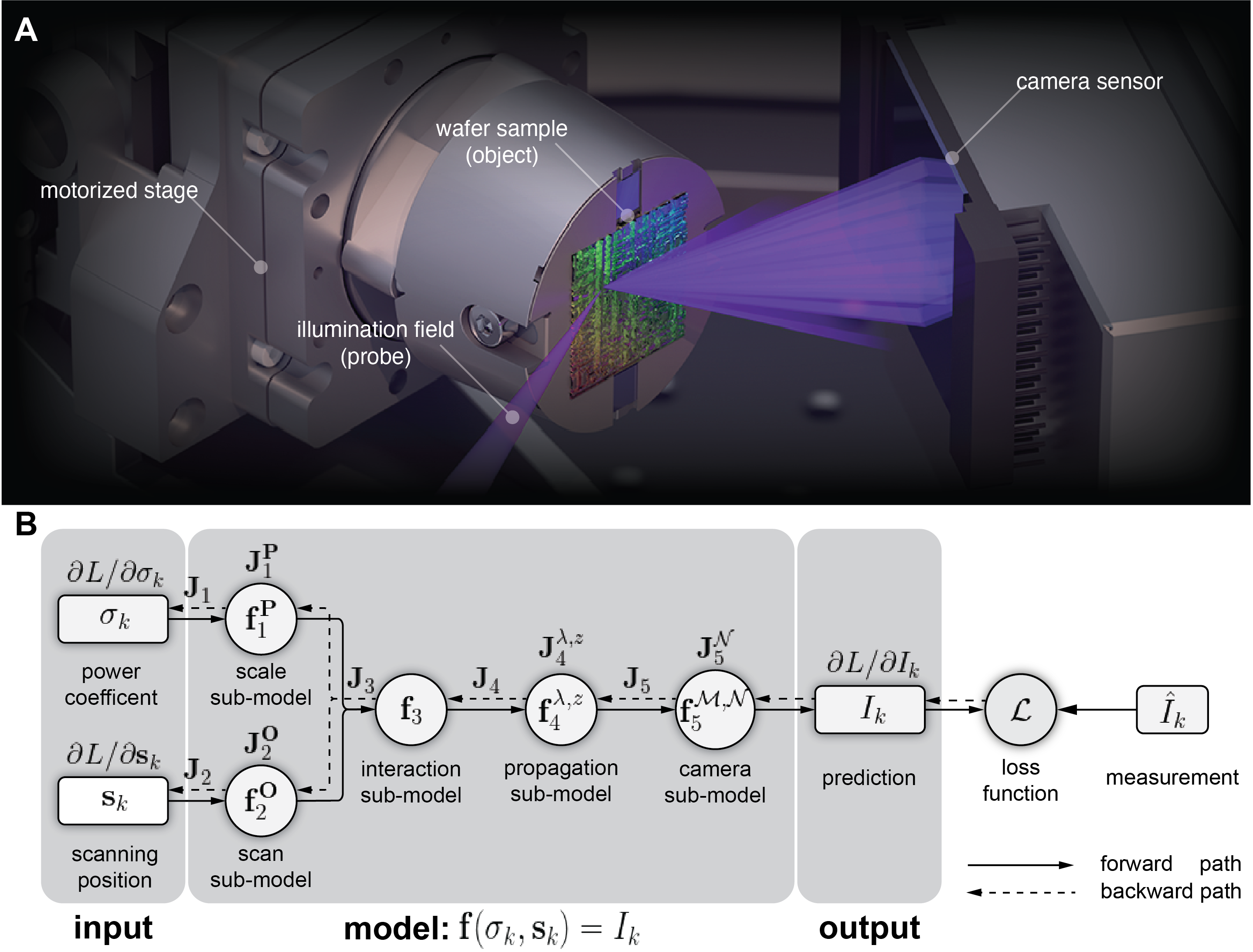}
  \caption{The ptychography process and the computational graph for the reconstruction. (\textbf{A}): Illustration of ptychography in reflection geometry at our EUV beamline. (\textbf{B}): The computational graph for the reconstruction. $\mathbf{f}$ and $\mathbf{J}$ denote the function and the corresponding Jacobian of each sub-model. The forward path (solid arrows) predicts the diffraction pattern measured by the camera and evaluates the loss function. The backward path (dashed arrows) computes the gradients with respect to the variables in each sub-model and the two inputs of the model in an accumulative manner.}
  \label{fig:computation_graph}
\end{figure}

It is customary to use a computational graph shown in Fig.~\ref{fig:computation_graph} (B) to describe the composite function $\mathbf{f}$, wherein the circles denote the functions in the composition, and the solid arrows indicate the composition relations. Ptychography can optimise all these variables ($\mathbf{s}_k$, $\sigma_k$, $\mathbf{P}$, $\mathbf{O}$, $\lambda$, $z$, and $\mathcal{N}$) by utilising the abundant and partially redundant information resulting from the partially overlapping illumination areas on the sample at adjacent scanning positions. 

With AD, computing the gradients of all variables becomes one unified process. AD essentially applies the chain rule to the composite function $\mathbf{f}$ and accumulates the Jacobians (matrices of partial derivatives) at all nodes when traversing the computational graph from output to input during a reverse pass. Conventional approaches can only achieve, in the best scenario, an efficiency equivalent to that of AD \cite{griewank2008evaluating,kakade2018provably}. Introducing regularisations to the optimisation is also straightforward by adapting the loss function. In each iteration, the gradients of variables will be modified by the regularisations accordingly. The strategy is described in detail in the Supplementary Information.

\subsection{Experimental setup and ptychography reconstruction}

In the experiment, the sample is illuminated with an EUV beam focused at an angle of incidence of 70 degrees relative to the sample's surface normal. We scan the illumination beam over a patterned sample consisting of 20-nm high gold-titanium structures deposited onto a silicon substrate. We acquire intensity measurements of the diffraction patterns at 255 scanning positions on a randomly perturbed regular grid. As we illuminate at grazing incidence angle, the probe is severely elongated along the direction of incidence with a size of approximately 30-by-100 \textmu m, resulting in an average scanning interval of 10 \textmu m along the elongated direction of the probe and 3 \textmu m along the other direction.

In reflection geometry, we must correct the distortion of the diffraction pattern due to sample tilt with respect to the optical axis. The tilt correction assumes that the camera is placed perpendicular to the direction of the specular reflection and requires information about both the propagation distance (88 mm measured from the sample to the camera) and the tilt angle (equal to the angle of incidence). Prior to the tilt correction, we subtract the noisy camera background, measured in the absence of EUV illumination, and align the zeroth order with the origin of the camera coordinate at the centre of the field-of-view (FOV). 

The EUV camera has 2048-by-2048 pixels with 15-by-15 \textmu m pixel size. Through the tilt correction, each diffraction pattern is interpolated non-uniformly onto a sampling grid that has 2002-by-3038 pixels with 15-by-5 \textmu m pixel size \cite{zhang2015high, porter2017general, Matsushima2020}. This results in a reduced FOV on the camera and, accordingly, a degraded resolution in the sample, both along the sample tilt direction. The sampling along the other direction remains unchanged. For EUV illumination at 18 nm wavelength, we achieve a sample resolution of 50-by-100 nm with a detection numerical aperture (NA) of 0.170-by-0.086. 

The tilt correction also generates a binary mask to separate the pixels inside and outside the reduced FOV. We set the higher and lower threshold of the pixel values to be $2^{16}-1$ and $4$, respectively, to discard the saturated and noise-corrupted pixels. The default initial value for the common background in all diffraction patterns to be reconstructed is a uniform value of one. We use a constant value of one as the initial value of the power coefficient at each scanning position. The reconstruction always starts with elliptical apertures as the initial guesses of the probe. The elliptical aperture's centre, size (long and short axes), and rotation angle can be configured individually for each wavelength. The initial object is a uniform value of one, representing a flat, unpatterned wafer sample. 

In the proof-of-principle experiment, we incorporate two harmonics of the HHG beam in forming the diffraction pattern. We achieve this by tilting the pair of wavelength-selective periscope mirrors to induce a shift of the filtering window in the spectral domain. However, as none of these two harmonics is aligned with the peak of the narrow-band filtering window, the beam intensity decreases, increasing the acquisition time from a few hundred milliseconds to tens of seconds per diffraction pattern. This can be avoided by using a pair of broadband periscope mirrors. The full scan of 225 positions was completed in 45 minutes, including 1 additional second per position to stabilise the motion stage and another 1 second to read out the image at a rate of 4 MHz using a single read-out port from the EUV camera, 

The wavelength-dependent behaviour of the sample in response to the illumination field necessitates the assumption of a dispersive sample in the reconstruction. Eventually, we reconstruct six illumination modes and one sample mode per wavelength to tackle the limited source spatial coherence and inherent instabilities. As each illumination mode and sample mode contains 12 million and 27 million parameters, we optimise over 200 million parameters with our universal ptychography algorithm based on the AD framework.


\clearpage
\section{Results}\label{sec2}

\subsection{Benchmarking with state-of-art open-source ptychography algorithm}

\begin{figure}[hbt]
  \includegraphics[width=\textwidth]{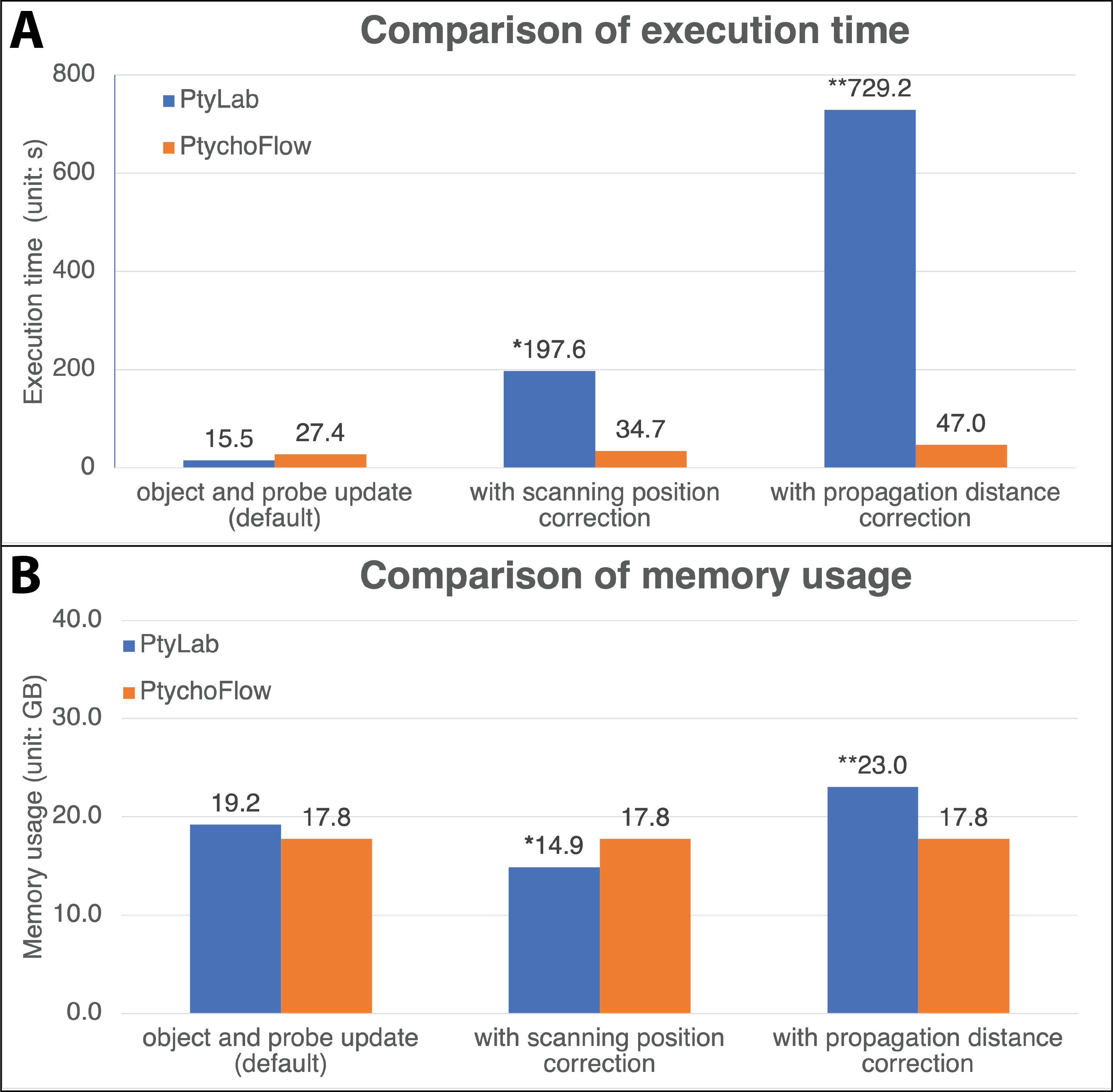}
  \caption{Performance comparison in execution time (\textbf{A}) and memory usage (\textbf{B}). 
  The comparison is conducted by reconstructing the experiment dataset with two wavelengths, and three illumination modes and one sample mode per wavelength. The numbers in the figures are averaged over five runs on a Nvidia RTX A600 GPU. * The number is obtained for the case of a single wavelength since PtyLab does not support scanning position correction in wavelength-multiplexed reconstruction. ** The number is adjusted since PtyLab only corrects the propagation distance per epoch instead of per batch.}
  \label{fig:computation_graph}
\end{figure}

We validate the benefit of the AD framework by conducting an objective performance comparison between our algorithm, named PtychoFlow, and other ptychography frameworks. Although several open-source code repositories like PtyPy \cite{enders2016computational}, PtychoShelves \cite{wakonig2020ptychoshelves}, and PtyLab \cite{loetgering2023ptylab} have been published, none offer the joint optimisation of an equivalent set of parameters discussed in this work. Among these algorithms, PtyLab is the most recent and, to the best of our knowledge, has implemented the most extensive list of features. It resembles important features such as the reconstruction of both multiple spectral harmonics and spatial modes, GPU acceleration, and momentum optimisers. However, it does not support switching on optimising correction features simultaneously. Instead, it alternates between updating different parameters in different iterations. We choose to compare our dataset's reconstruction in both PtychoFlow and PtyLab. We focus on objective goals such as execution time and memory usage. In contrast, identifying the suitable initialisations and hyperparameters for optimisation is subjective and will therefore not be pursued in this study.  

In Fig.~\ref{fig:computation_graph} ({A}), we compare the execution time per iteration across various scenarios using identical computational hardware. In the default setting of updating only the pixel values in the object and probe, PtyLab uses the mPIE engine \cite{maiden2017further} and outperforms PtychoFlow. It is worthwhile noting that the speed difference may be attributed to the different GPU acceleration packages used: CuPy \cite{cupy_learningsys2017} by PtyLab and TensorFlow by our algorithm, although both operate on the same Nvidia's CUDA platform. Additionally, PtyLab employs a two-step propagator for wavelength-multiplexed reconstructions \cite{loetgering2021tailoring}, while our propagator is based on the chirp z-transform (CZT). 

In subsequent comparisons on scanning position and propagation distance correction, we use the pcPIE \cite{maiden2012annealing} and zPIE \cite{loetgering2020zpie} engines in PtyLab. Here, the benefit of AD becomes apparent, as indicated by the drastic speed advantage. Like other conventional ptychography algorithms \cite{zhang2013translation, dwivedi2018lateral, ruan2022adaptive}, PtyLab updates the additional variables via a secondary optimisation loop alongside the primary iterative routine to reconstruct the probe and the object while updating all parameters in the AD framework is a unified process. We achieved this by using differentiable operations such as the Fourier transform shift theorem for the scanning and the CZT for the propagator. 

When comparing the usage of GPU's Video Random-Access Memory (VRAM), a critical resource for ptychographic reconstructions, we found in Fig.~\ref{fig:computation_graph} ({B}) that, thanks to AD, incorporating additional parameters does not impact the memory usage in PtychoFlow, while PtyLab tends to use extra VRAM for the correction. The overall performance comparison confirms that the computational complexity scales much more consistently and favourably for PtychoFlow than for conventional algorithms as used in PtyLab. 

\subsection{Wavelength-multiplexed diffractive imaging}
\label{sec: CDI}

\begin{figure}[htb]
  \includegraphics[width=\textwidth]{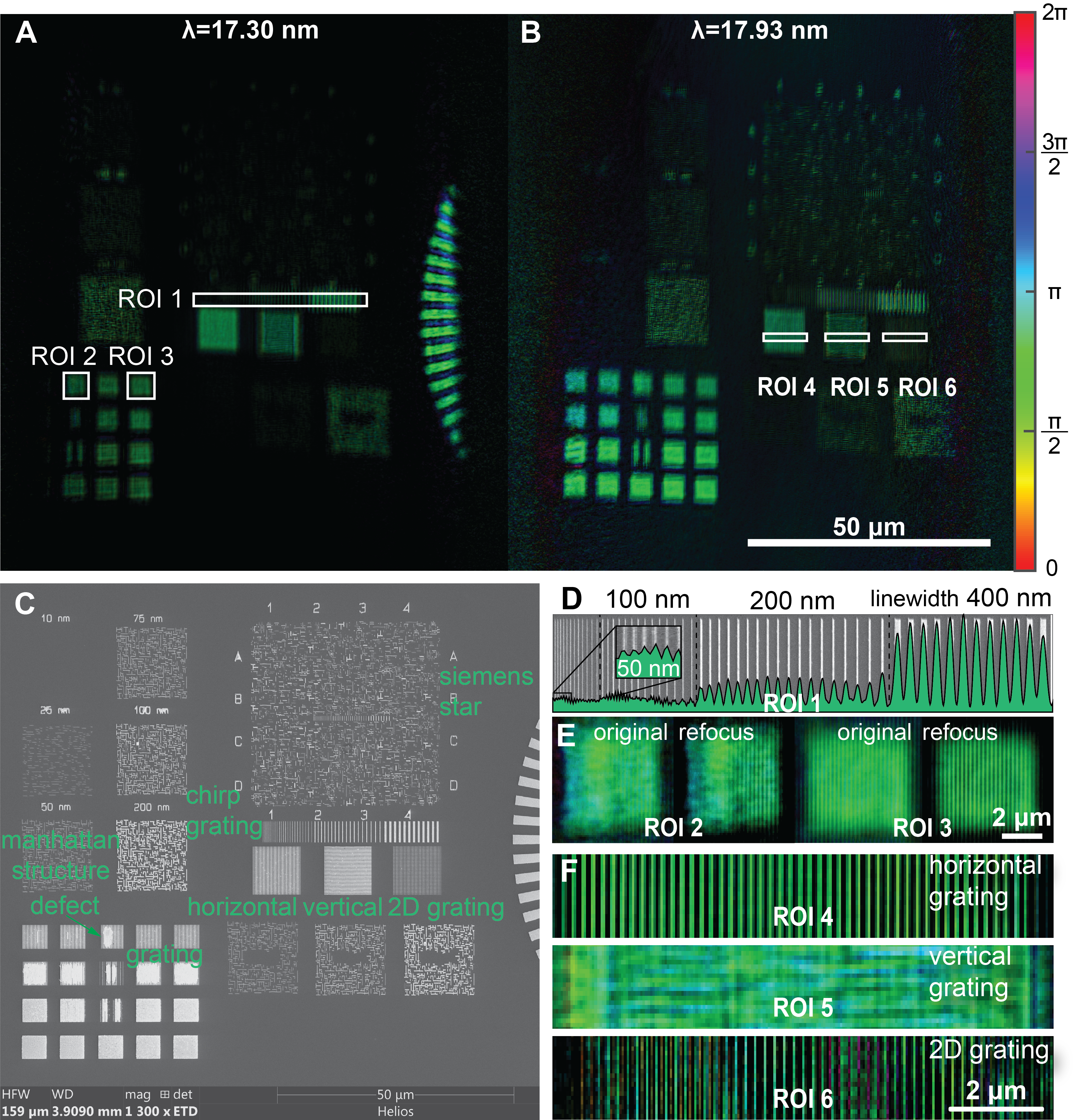}
  \caption{\textbf{Dual wavelengths nanostructure imaging.} (\textbf{A} and \textbf{B}): Ptychographic reconstructions of nanostructures on the wafer sample at wavelength 17.30 nm (A) and 17.93 nm (B). The brightness and hue of the image represent the amplitude and phase of the complex-valued object, respectively. The reconstructed objects show shifted field-of-views as the probes at different wavelengths illuminate different areas on the sample. (\textbf{C}): SEM image of the wafer sample with various types of structures and manufacturing defects for comparison. (\textbf{D}): Profile of the chirped grating in ROI 1 consisting of 200 nm, 100 nm, and 50 nm lines with varying spacing. Inset: the finest part of the chirped grating with 150 nm pitch and 50 nm linewidth. (\textbf{E}): Gratings in squared areas ROI 2 (with a manufacturing defect) and ROI 3, before (left) and after (right) refocusing for enhancing contrast. (\textbf{F}): Zoom-in of the reconstructed structures in ROI 4-6. Due to the non-uniform sample resolution in the reflection geometry, only the horizontal grating and the horizontal part of the 2D grating can be resolved.}
  \label{fig:nanostructure_imaging}
\end{figure}

The reconstructed objects presented in Fig.~\ref{fig:nanostructure_imaging} (A) and (B), corresponding to sample modes at wavelengths 17.30 nm and 17.93 nm, respectively, exhibit shifted field-of-views on the wafer sample. The cross-correlation analysis of these objects indicates a shift of about 30 \textmu m, mainly in the direction perpendicular to the direction of sample tilt and the elongation direction of the probes. This shift suggests that the probes at the different wavelengths illuminate different areas of the sample. A similar spatial dispersion effect of the HHG beam has been reported in \cite{zhang2016ptychographic}.

For a distance of 160 mm between the focusing optics and the sample, we found that the propagation directions of the two harmonics deviate from each other only by a negligible angle of 0.03 mrad at the pupil (see Supplementary Information Sec.~10). Additionally, noticeable non-uniform deformation in each reconstructed object can be observed, which can most likely be attributed to the errors in the tilt plane correction procedure. However, the cause of the difference in the deformations at different wavelengths is a subject of future investigation. 

In Fig.~\ref{fig:nanostructure_imaging} (D), the chirped grating profile in region-of-interest (ROI) 1 is displayed. The profile is obtained by summing the intensity of the reconstructed object along the direction parallel to the grating lines. From left to right, the grating consists of equally spaced 400 nm lines (800 nm pitch), 200 nm lines with decreasing pitches from 800 nm to 600 nm, 100 nm lines with decreasing pitches from 600 nm to 350 nm, and 50 nm lines with decreasing pitches 350 nm to 150 nm. The inset plot on the leftmost side shows the finest part of the grating, in which the peaks align well with the locations of the 50 nm lines in the background SEM image. The obtained resolution of 50 nm is the diffraction limit with a NA of 0.17 at 18 nm wavelength.  

To improve the contrast of the reconstructed object, we employ a refocusing approach by propagating the entire object to a series of planes between -25 \textmu m and +25 \textmu m with an interval of 200 nm using the angular spectrum method. Specially, we focus on refining the gratings in ROI 2 and ROI 3, both with 200 nm pitch and 50\% filling ratio. By extracting the strength of the spatial frequency component at $5\ \text{\textmu m}^{-1}$ from the Fourier transform of the grating profile in each plane, we can obtain a curve as a function of the z location with fast oscillations due to the Talbot effect and a slowly varying envelope due to defocusing of non-periodic features in the periodic grating. By locating the maximum of this curve, we can determine a defocus distance of -3.51 \text{\textmu m} and -3.31 \text{\textmu m} for ROI 2 and ROI 3, respectively. See Supplementary Information for details.

For ptychography in the reflection geometry, the resolution of the reconstructed object is compromised in the direction of the sample tilt and the elongation direction of the probe. This compromise is illustrated in Fig.~\ref{fig:nanostructure_imaging} (F) by the reconstructed object with a non-symmetric resolution of 50 nm in the horizontal direction and 100 nm in the vertical direction. As a result, the horizontal grating (ROI 4) exhibits distinct periodic structures with a pitch of 200 nm and a linewidth of 100 nm, while the vertical grating (ROI 5) with identical pitch and linewidth appears almost uniform. In the 2-dimensional grating (ROI 6), the horizontal components are resolved as a 1-dimensional grating, while the vertical components are effectively indistinguishable. We validate the resolution via spatial frequency analysis and Fourier ring correlation (FRC) \cite{van1982arthropod,koho2019fourier} (see Supplementary Information). In FRC, we correlate the reconstructed objects with the SEM images to avoid some possible artefacts of the usual FRC. Both confirm that a spatial resolution between 100-200 nm has been achieved for this specific case. 

\subsection{3D structure information characterisation}

\begin{figure}[hbt]
  \includegraphics[width=\textwidth]{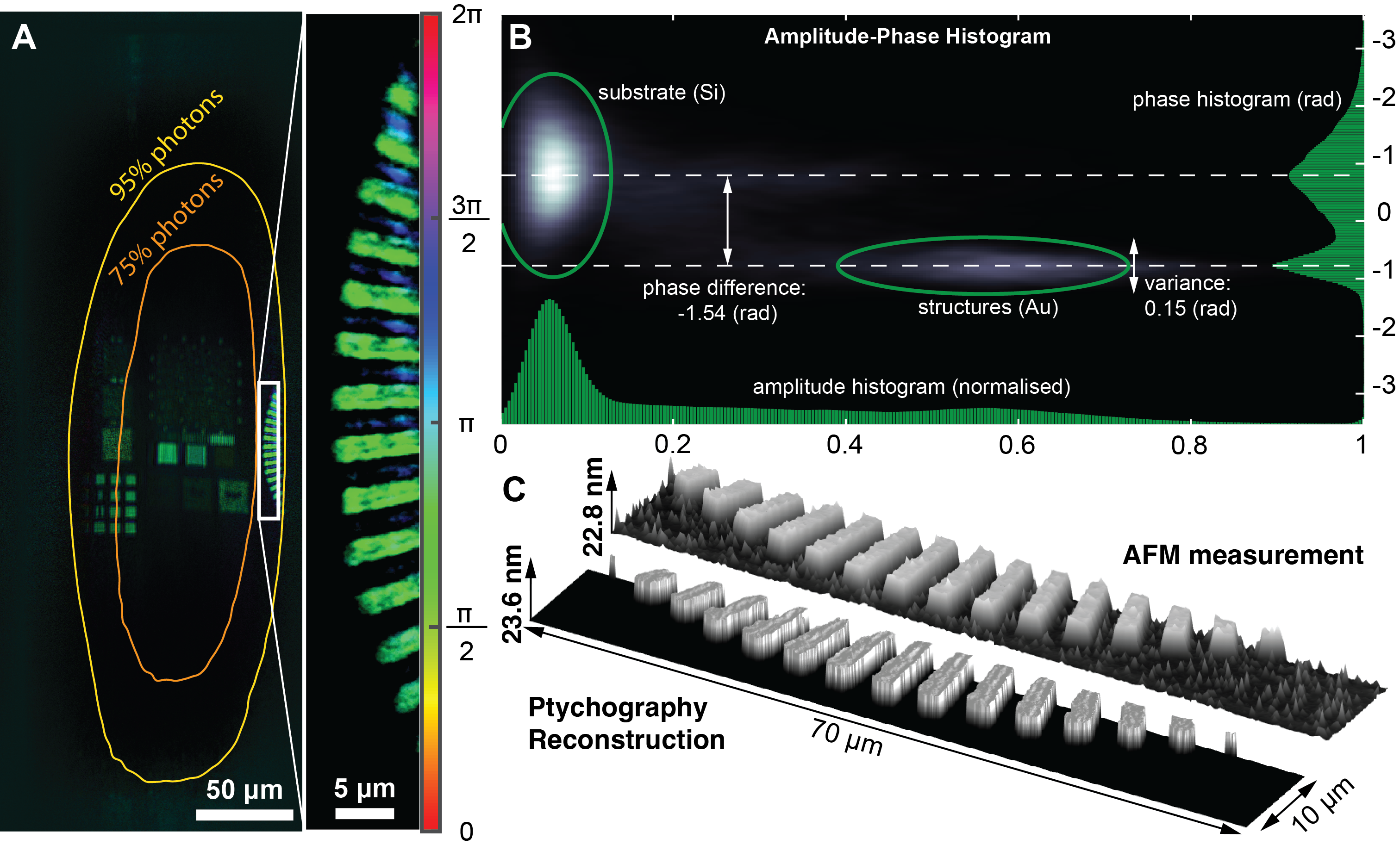}
  \caption{3D structure information characterisation. (\textbf{A}): The entire field-of-view of the reconstructed complex-valued object. The ellipsoidal contours in orange and yellow mark the areas illuminated with 75\% and 95\% of the total photons during the scanning process, respectively. Inset: the reconstructed Siemens star with 90 spokes and a radius of 60 \textmu m. (\textbf{B}): Distribution of pixel values in amplitude (normalised) and phase coordinates. The green circles indicate the clusters of pixels for the Si substrate and Au structures, respectively. (\textbf{C}): 3D representation of the AFM measurement and the Siemens star reconstructed by ptychography. The height is computed using the phase difference determined in (B), the incident angle, the wavelength, and with prior knowledge of the materials. }
  \label{fig:3D_characterisation}
\end{figure}

In Fig.~\ref{fig:3D_characterisation} (A), the contours indicate the regions where specific percentages of the total photons illuminate the sample during the scanning process. Consider the part of the Siemens star reconstructed in the full field-of-view located in between the 75\% and 95\% contours. This ring area is illuminated by only 25\% of the total amount of photons, and only a small fraction of this amount of photons is scattered by the structures, contributing to the formation of the diffraction pattern, while almost no photon is scattered by the substrate. 

The reconstruction of the complex-valued object enables the segmentation of the sample into the gold structures and the silicon substrate based on the complex values and spatial locations of the pixels. In Fig.~\ref{fig:3D_characterisation} (B), we present the amplitude-phase histogram of the pixels in the reconstructed Siemens star, where two clusters of pixels can be identified: one with higher amplitudes representing the structures and the other with lower amplitudes representing the substrate. For illumination at 17.30 nm wavelength with a 70 degrees incidence angle, the reflectivities of the gold and the silicon 0.37 and 0.0077, respectively \cite{henke1993x}. The structures reveal small phase variations, represented by a uniform green colour in the reconstructed Siemens star. The substrate, however, exhibits a large phase variation due to the low amplitude. This emphasises the need for caution when applying L1 and TV regularisations to the amplitude as both tend to drive the low amplitude further to zero and thus can introduce extra uncertainties when determining the structure information.

By averaging the phase over all pixels in each cluster, the phase difference between the structures and the substrate can be computed. This phase difference arises from two sources: (1) the phase difference induced by the reflection at different interfaces through the Fresnel relations, and (2) the phase difference accumulated through propagation over different distances. With the prior knowledge of the respective compositions of gold and silicon in the structures and substrate, the first part can be obtained by computing the phase difference between the Fresnel coefficient at the surfaces. The second part, which is directly proportional to the height of the structures with respect to the substrate, can be obtained by subtracting the first part from the reconstructed phase difference. In this work, the thus estimated height of $23.6\pm0.62$ nm agrees well with both the 20 nm nominal design value and the height retrieved from the AFM measurement of $22.8\pm1.45$ nm (see Supplementary Information). Fig.~\ref{fig:3D_characterisation} (C) shows a 3D representation of the segmented Siemens star with the values of the pixels belonging to the two clusters, the substrate and the structures, set to zero and actual height, respectively.

\subsection{EUV focusing optics characterisation}

\begin{figure}[hbt]
  \includegraphics[width=\textwidth]{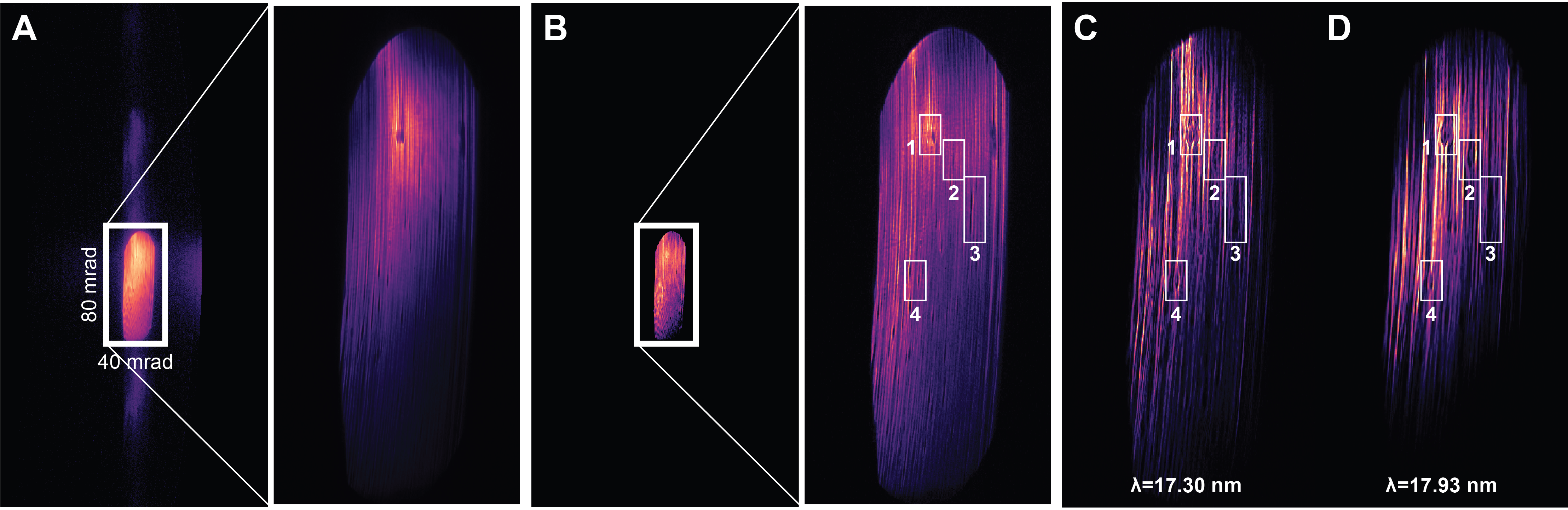}
  \caption{EUV focusing optics characterisation. (\textbf{A}): Sum of the measured diffraction patterns in logarithmic scale with zeroth order in the shape of the ellipsoidal focusing mirror and higher orders cropped in the horizontal direction due to the sample tilt in reflection geometry. Inset: zoom-in of the zeroth order in linear scale. (\textbf{B}): Reconstructed common background intensity in logarithmic with a clearly distinct ellipsoidal shape representing the pupil of the illumination system. Inset: zoom-in of pupil in linear scale. (\textbf{C} and \textbf{D}): Reconstructed pupil functions (Fourier transform of the respective reconstructed illumination probes) at wavelengths 17.30 nm (C) and 17.93 nm (D). The rectangularly shaped ROI 1-4 show the locations of the defects on the EUV focusing mirror.}
  \label{fig:EUV_focusing_optics}
\end{figure}
Fig.~\ref{fig:EUV_focusing_optics} (A) and (B) show the sum of all diffraction patterns and the common background, respectively. Under the first Born approximation with the regular multiplicative assumption made in ptychography, the far-field diffraction pattern can be interpreted as the Fourier transform of the object convolved with the field distribution in the pupil when using a focused beam to illuminate the sample. Because of this convolution relation, both the zeroth diffraction order, which is in the shape of the ellipsoidal focusing mirror, and higher diffraction orders are visible in Fig.~\ref{fig:EUV_focusing_optics} (A). In the presence of specular reflection by the sample, the Fourier transform of this uniform signal is a delta function in the far-field. As a result of convolving with the delta function, the common background on the camera is almost identical to the pupil plane field distribution despite the minor contributions by other effects like stray light due to spurious reflections in the beamline (see Supplementary Information for more details). Therefore, the common background can be used to characterise the focusing optics. Fig.~\ref{fig:EUV_focusing_optics} (B) clearly reveals grooves of grinding tool marks and defects on the surface of the ellipsoidal focusing mirror.

The pupil function can also be obtained by applying the inverse Fourier transform to the reconstructed probe. By comparing the common background with the two pupil functions at two wavelengths shown in Fig.~\ref{fig:EUV_focusing_optics} (C) and (D), we can confirm the presence of the grinding tool marks and the defects on the focusing optics. ROI 1-4 reveal that the locations of the defects are well aligned in both the pupil functions and the common background, yet with distinguishable visual effects. For instance, those defects in ROI 2 \& 4 are more discernible in the pupil functions than in the common background. It is intriguing to note that the different harmonics illuminate different areas on the focusing mirror. This results in a deviation between the propagation directions of the probes, leading to relatively shifted field-of-views of the reconstructed objects as mentioned in Sec.~\ref{sec: CDI}. 

\clearpage
\section{Discussions and Conclusions}\label{sec3}

The introduced universal ptychography algorithm based on the automatic-differentiation (AD) framework has been demonstrated to be a powerful tool for EUV diffractive imaging that is crucial for future semiconductor metrology applications. It covers both transmission and reflection (through tilt correction) configurations and can deal with source instabilities and numerous sources of experimental uncertainties. 

The AD framework allows us to jointly optimise the high-dimensional arrays of the probe and the object, along with other variables involved in all relevant physical processes. As long as a physical process can be mathematically described using differentiable operations, it can be incorporated into the universal ptychography model presented in the Introduction, and its variables can be retrieved using AD and gradient-descent type optimisation. We demonstrate through a comparison with a state-of-art open-source ptychography algorithm that the AD approach is superior in both execution time and memory usage. Combined with the modular design, our versatile and expandable ptychography algorithm is capable of handling diverse challenges and is future-proof to meet the requirements of next-generation metrology applications.

Implementing the algorithm on mainstream machine learning platforms like TensorFlow provides GPU acceleration as an additional benefit. However, the limited memory of the GPU also imposes a constraint on the model's size. In ptychography, computing the Fourier transform for field propagation is the most time-consuming operation. Because ptychography computes one field propagation per spatial mode per spectral harmonic, computation time increases rapidly as the number of spectral harmonics and spatial modes increases. This issue can be mitigated by distributed optimisation on GPU clusters.

In our case, reconstructing the probe and the dispersive object with six illumination modes and one sample mode per wavelength, and with a total of two wavelengths, requires twelve propagations of the fields, consuming 97.58\% of the total 48 GigaBytes memory on an Nvidia RTX A6000 GPU. A considerable portion of this memory consumption, 7.42 GigaBytes, is on storing the dataset to avoid the time-consuming task of constantly passing the data back and forth between the GPU and CPU. 

To achieve the results shown in this paper, we optimise a ptychography model containing variables with a total number of 201 million parameters to retrieve information from a dataset containing frames with a total number of 1.1 billion pixels. The 1:5 ratio between the numbers of the parameters and the pixels suggests there is still a sufficient abundance of information available in the data. To gain some insights, our ptychography model contains a comparable number of parameters as some famous deep neural network models for image/language processing, such as AlexNet \cite{krizhevsky2017imagenet} (60 million), GPT-1 \cite{radford2018improving} (117 million), VGGNet \cite{simonyan2014very} (134 million), and Bert \cite{devlin2018bert}  (340 million). 

Among these parameters, the illumination modes, required to deal with the lack of spatial coherence of the HHG EUV probe, can be decreased by increasing the source stability. This reduces both the computation time per iteration and the number of iterations. Besides, by pre-calibrating the probe and other experimental parameters, leaving the object as the only variable, one can significantly reduce the number of parameters and hence further achieve fast reconstructions.

\begin{figure}[hbt]
\centering
  \includegraphics[width=0.8\textwidth]{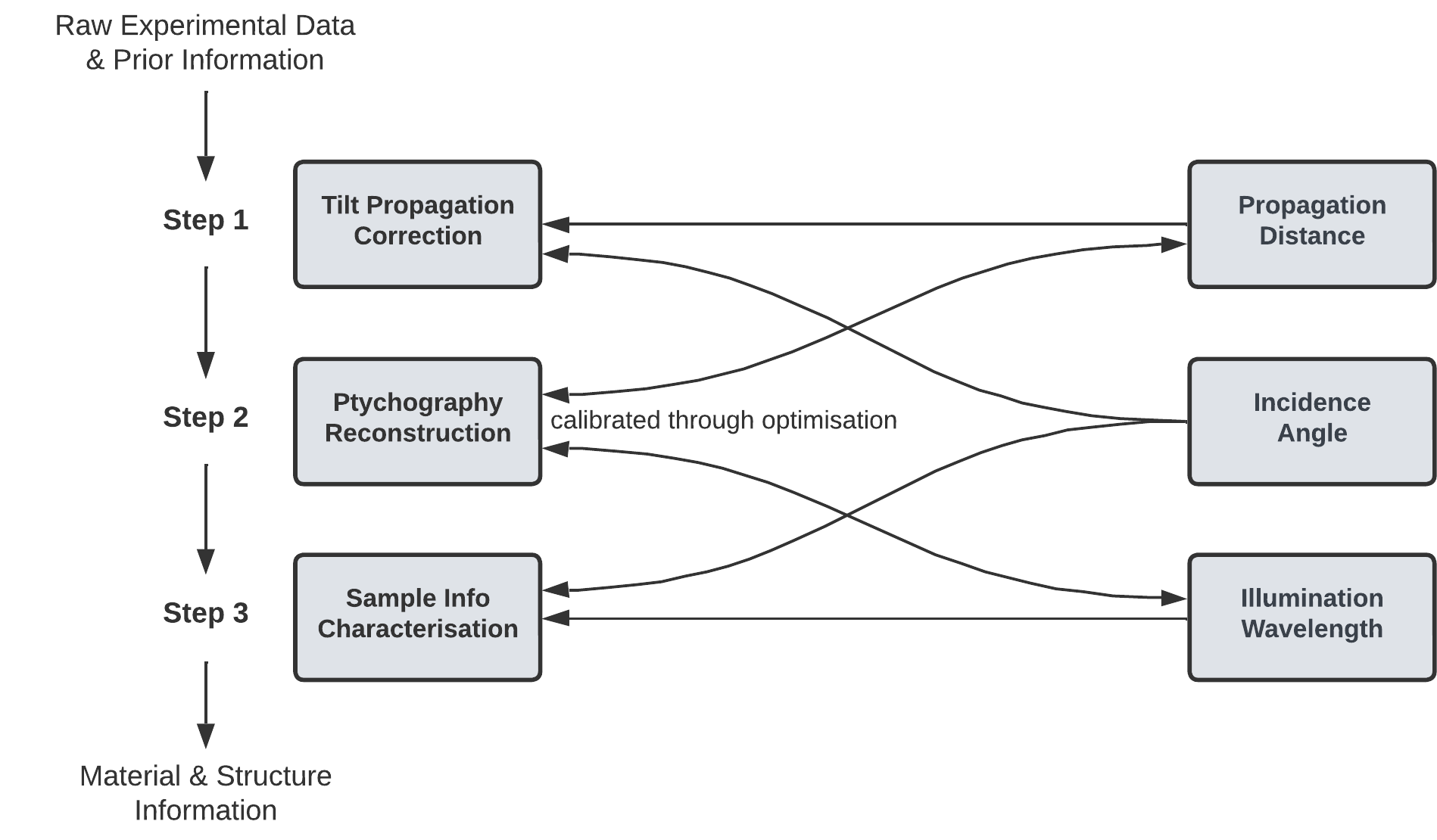}
  \caption{Intertwined relations between the main steps for quantitative ptychographic imaging and the relevant variables. }
  \label{fig:step_variable_relation}
\end{figure}

Finally, achieving unique and unambiguous results is crucial for quantitative imaging approaches with metrology applications at the industrial level. Fig.~\ref{fig:step_variable_relation} depicts the intertwined relations between the three main steps for quantitative imaging and the three relevant variables.  This necessitates incorporating the tilted propagation into the model alongside the ptychography process. For future research, calibrating these three variables through optimisation, along with the reconstruction of the probe and the object, is critical to secure the unique and unambiguous characterisation of the sample information.

In conclusion, our AD-based ptychography algorithm successfully reconstructs images of the samples using data acquired with multiple harmonic frequencies and limited spatial coherence by considering multiple illumination and sample modes per wavelength. AD is essential for managing complex, large-scale ptychography models and allows the algorithm to account for various uncertainties arising during the experiments. The wavelength-multiplexed reconstruction using HHG EUV illumination in reflection geometry demonstrates the possibility of determining both material and structure information for wafer samples. These findings reveal the capabilities of our ptychography algorithm to release the full potential of advanced HHG sources in view of EUV diffractive imaging and establish ptychography as an imaging metrology approach with the potential for high-resolution and high-throughput applications.
%
%
\clearpage
\section{Methods and Materials}\label{sec4}
\subsection{EUV beamline for diffractive imaging experiments}

\begin{figure}[hbt]
 \includegraphics[width=\textwidth]{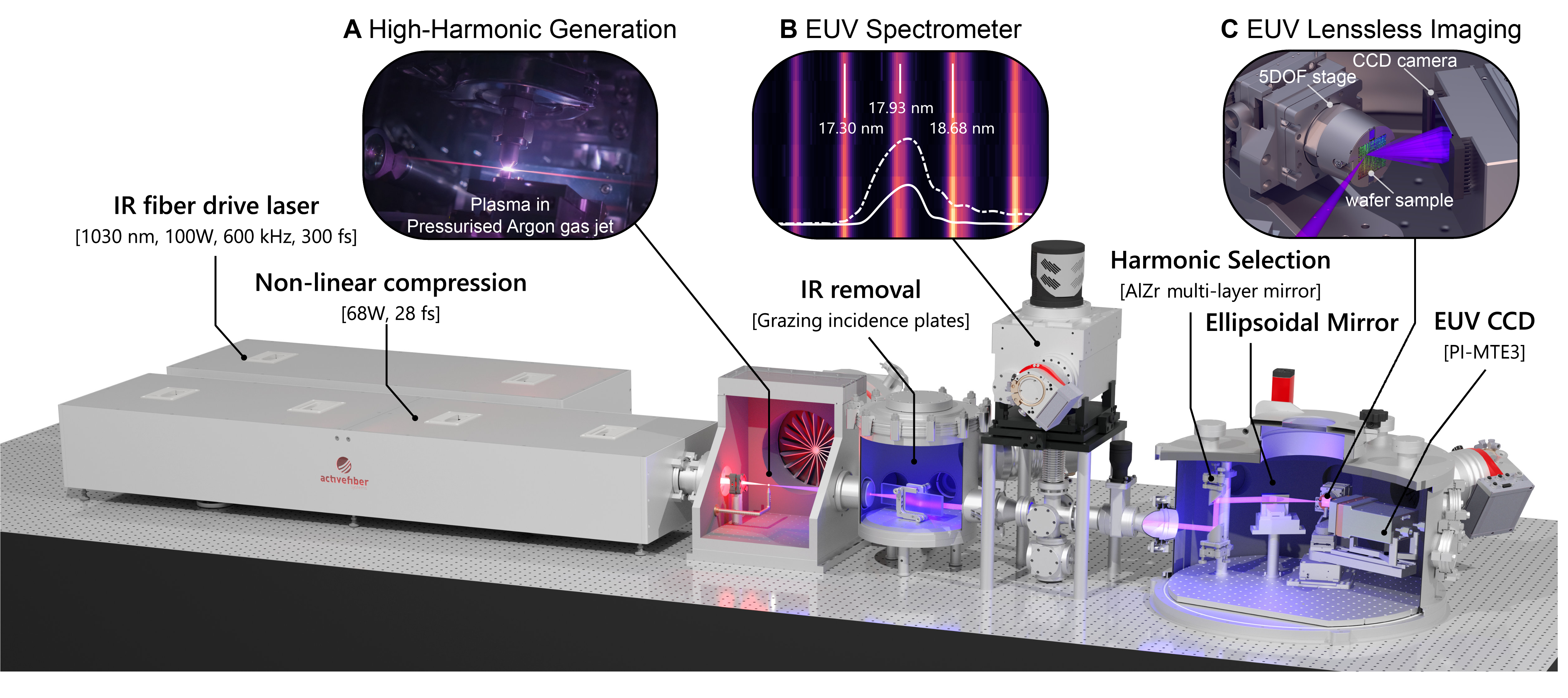}
 \caption{Higher harmonics are generated in a pressurised Argon gas jet in the vicinity of the focus of an infrared pump laser, as shown by inset (\textbf{A}), and are separated from the infrared pump laser. The HHG spectrum is calibrated using a transmission spectrometer with high-density gratings. A single harmonic can be selected with spectral filtering using a pair of Al-Zr mirrors, whose spectrum is depicted by the white dashed (one mirror) and solid (two mirrors) curves in inset (\textbf{B}). The illumination beam is focused onto a patterned sample by an ellipsoidal mirror, and the resulting diffraction pattern is captured by a CCD camera sensitive to EUV photons for ptychography reconstruction.}
 \label{fig:setup}
\end{figure}

In the experiment, the compact HHG source is pumped by a Ytterbium-doped fibre laser (Active Fibre System UFFL 100) with a centre wavelength at 1030 nm. The pump laser emitted 167 \textmu J pulses with a pulse length of 300 fs, achieving an average power of 100 W at a repetition rate of 600 kHz. These pulses are further compressed to a pulse length of 28 fs by a Krypton-filled hollow core fibre (HCF) and a set of chirped mirrors. Following compression, the pulsed infrared (IR) beam remains with 114 \textmu J pulse energy and 68 W average power. 

The IR pump laser is focused into a 10-bar pressurised Argon gas jet with a focal spot size of 25.6 \textmu m, placed between a 300 \textmu m sized gas nozzle and a gas catcher as illustrated by Fig.~\ref{fig:setup} (A). This enables the generation of higher harmonics up to the soft X-ray regime, with the strongest harmonic in the EUV range obtained at 68 eV photon energy, or 18 nm wavelength, with a photon flux of 2 $\cdot$ $10^{11}$ photons per second at the gas jet within 1\% of the bandwidth as reported in previous work \cite{tschernajew2020high}. By using Neon gas, our HHG source is capable of generating EUV radiation centred at 13.5 nm. 

Due to the significant power disparity (eight orders of magnitude) between the generated weak higher harmonics and the much stronger IR pump laser, several measures were implemented to separate the two. Firstly, an annular dichroic mirror with an aperture of 2 mm rejects approximately half of the IR power by utilising the larger divergence of the IR beam relative to that of the higher harmonics. Secondly, a pair of grazing incidence plates (GIP), positioned at the Brewster angle for the IR beam, further attenuate the IR beam while still reflecting the higher harmonics with a total efficiency of 54\%. Finally, a pair of 200 nm thick aluminium freestanding foils, which acts as a long-pass filter with a transmission cut-off of 73 eV (equivalent to 17 nm) for the higher harmonics, block the remaining IR beam.
 
In order to calibrate the spectrum of the broadband EUV beam, a gold-coated flip mirror is inserted into the beamline to redirect the beam towards a spectrometer before selecting a single harmonic for illumination. The spectrometer's working principle is explained in detail in \cite{goh2015fabrication}. A portion of the calibrated spectrum, ranging from 16.5 nm to 19.5 nm, is presented in Fig.~\ref{fig:setup} (B).

The selection of a single harmonic is achieved using spectral filtering via a pair of consecutive multi-layer mirrors (MLMs) arranged in a periscope-like configuration. These mirrors are positioned parallel at a 45-degree angle with respect to the optical axis and coated with alternating layers of zirconium (Zr) and aluminium (Al). The spectral filter exhibit a reflectivity peak at the centre wavelength of 18 nm with a combined peak reflectivity of 21\% and a full width at half maximum (FWHM) of 0.6 nm. The white solid curve in Fig.~\ref{fig:setup} (B) illustrates the reflection spectrum of the Zr-Al mirror pair.

The selected single harmonic incident at 10 degrees grazing angle relative to the surface on a Ruthenium-coated ellipsoidal mirror was focused onto the sample, forming a probe of 33 \textmu m by 34 \textmu m in the focal region. The size of the probe is confirmed through a knife-edge measurement in the plane perpendicular to the probe. However, due to the tilt of the sample for 20 degrees grazing incidence, the probe is elongated in the tilt direction by a factor of three.

To acquire data for ptychography reconstructions, a set of linear slip-stick piezo stages is used to scan the probe over the sample surface. Additionally, three more slip-stick piezo stages are employed to align the sample's depth, azimuth angle, and incidence angle, as shown in Fig.~\ref{fig:setup} (B). The entire assembly is stabilised with interferometry-based closed-loop configurations.

The beam reflected by the sample propagates freely towards a full-vacuum CCD camera (PI-MTE3 2048) with 2048-by-2048 pixels and a pixel size of 15-by-15 \textmu m. The camera is mounted on a rotation stage to ensure a normal incidence of the specularly reflected beam.

\subsection{Algorithm implementation}
We have implemented our algorithm to achieve the results in this work as a monolithic and modular package, called \textit{PtychoFlow}. The package is developed in the Python environment with its core functionalities - including the physical model for ptychography (described in the Introduction) and the regularised optimisation loop (described in the Supplementary Information) - implemented on Tensorflow, a mainstream open-source machine learning platform, and the peripheral functions, such as data I/O, data processing and visualisation, implemented using popular Python open-source libraries such as Numpy, Matplotlib, etc.

\section{Acknowledgement}

This publication is part of the project Lensless Imaging of 3D Nanostructures with Soft X-Rays (LINX) with project number P16-08 of the Perspectif research programme financed by the Dutch Research Council (NWO).

The authors acknowledge Mengqi Du, Antonios Pelekanidis, and Prof. Stefan Witte for providing consultancy and testing datasets during the development of our algorithm. We also thank our dedicated technician, Roland Horsten, for his important work on hardware and software. 

\bibliography{sn-bibliography}


\end{document}